\documentclass[preprint,12pt]{elsarticle}

\usepackage{amssymb}
\usepackage{bm}
\usepackage{graphicx}
\usepackage{dcolumn}
\usepackage{color}

\newcommand{\beq}{\begin{equation}}
\newcommand{\eeq}{\end{equation}}
\newcommand{\beqn}{\begin{eqnarray}}
\newcommand{\eeqn}{\end{eqnarray}}
\journal{Physics Letter B}
\begin{document}

\begin{frontmatter}

\title{Novel structure for magnetic rotation bands in $^{60}\rm Ni$}

\author[PKU]{P.W. Zhao}

\author[PKU]{S.Q. Zhang}

\author[BNU]{J. Peng}

\author[PKU]{H.Z. Liang}

\author[TMU,PKU]{P. Ring}

\author[BUAA,PKU,Stell]{J. Meng\corref{corr1} } \ead{mengj@pku.edu.cn}

\cortext[corr1]{Corresponding author}
\address[PKU]{State Key Laboratory of Nuclear Physics and Technology, School of Physics, Peking University, Beijing 100871, China}
\address[BNU]{Department of Physics, Beijing Normal University, Beijing 100875, China}
\address[TMU]{Physik Department, Technische Universit\"at M\"unchen, D-85747 Garching, Germany}
\address[BUAA]{School of Physics and Nuclear Energy Engineering, Beihang University, Beijing 100191, China}
\address[Stell]{Department of Physics, University of Stellenbosch, Stellenbosch, South Africa}

\begin{abstract}
The self-consistent tilted axis cranking relativistic mean-field theory based on a point-coupling interaction has been established and applied to investigate systematically the newly observed shears bands in $^{60}\rm Ni$. The tilted angles,  deformation parameters, energy spectra, and reduced $M1$ and $E2$ transition probabilities have been studied in a fully microscopic and self-consistent way for various configurations and rotational frequencies.
It is found the competition between the configurations and the transitions from the magnetic to the electric rotations have to be considered in order to reproduce the energy spectra as well as the band crossing phenomena.
The tendency of the experimental electromagnetic transition ratios $B(M1)/B(E2)$ is in a good agreement with the data, in particular, the $B(M1)$ values decrease with increasing spin as expected for the shears mechanism, whose characteristics are discussed in detail by investigating the various contributions to the total angular momentum as well.
\end{abstract}



\end{frontmatter}


\section{Introduction}

The study of bands with high angular momenta has been at the forefront of nuclear structure physics for several decades. Many exciting phenomena have been discovered and predicted in this field such as backbending~\cite{Johnson1971Phys.Lett.}, alignment phenomena~\cite{Stephens1972Nucl.Phys.,Banerjee1973Nucl.Phys.}, superdeformed rotational bands~\cite{Twin1986Phys.Rev.Lett.}, magnetic rotation~\cite{Frauendorf1994,Frauendorf1997Z.Phys.A} and chiral phenomena~\cite{Frauendorf1997NP}.
In particular, magnetic rotation has become one of the most important phenomena in the last decade.

The term ``magnetic rotation'' has been attributed to the so-called
shears bands, which have strong $M1$ transitions and very weak $E2$
transitions. The explanation of such shears bands in terms of the
shears mechanism was firstly given in
Ref.~\cite{Frauendorf1993Nucl.Phys.}. In shears bands, the magnetic
dipole vector, which arises from proton particles (holes) and
neutron holes (particles) in high-$j$ orbitals, rotates around the
total angular momentum vector. At the bandhead, the proton and
neutron angular momenta are almost perpendicular to each other.
Along the bands, energy and angular momentum are increased by a
step-by-step alignment of the proton and neutron angular momenta
along the total angular momentum. Consequently, the direction of the
total angular momentum in the intrinsic frame does not change so
much and regular rotational bands are formed in spite of the fact
that the density distribution of the nucleus is almost spherical or
only weakly deformed. In order to distinguish this kind of rotation
from the usual collective rotation in well-deformed nuclei (called
electric rotation), the name "magnetic rotation" was introduced in
Ref.~\cite{Frauendorf1994}, and also discussed in
Ref.~\cite{Frauendorf1997Z.Phys.A}.

From the experimental side, long cascades of magnetic dipole $\gamma$ ray transitions were observed firstly in the neutron deficient Pb nuclei in the early 1990s~\cite{Clark1992Phys.Lett.,Baldsiefen1992Phys.Lett.,Kuhnert1992Phys.Rev.C}. Later, the lifetime measurements by Clark \textit{et al.} for four $M1$ bands in $^{198,199}\rm Pb$ provided a clear evidence for magnetic rotation~\cite{Clark1997Phys.Rev.Lett.}. From then on, more and more magnetic rotation bands have been observed not only in the mass region of $A\sim190$ but also in $A\sim80$, $A\sim110$, and $A\sim140$ regions. To date, more than 130 magnetic dipole bands have been identified in these four mass regions~\cite{Amita2000AtomicDataandNuclearDataTables}. In a recent experiment, magnetic rotation bands have also been observed in $^{60}\rm Ni$~\cite{Torres2008Phys.Rev.C}, which makes $^{60}\rm Ni$ the lightest system with magnetic rotation observed so far.

From the theoretical side, magnetic rotation bands have been well explained by spherical shell-model calculations~\cite{Frauendorf1996Nucl.Phys.} and cranking mean-field models~\cite{Frauendorf1993Nucl.Phys.}. In particular, the mean-field approaches have been widely used because here it is easy to construct classical vector diagrams showing
the angular momentum composition which is of great help in representing the structure of these rotational bands. It should be noticed that for magnetic rotations the axis of the uniform rotation does not coincide with any of the principal axis of the density distribution. Therefore, a description of these bands requires a model going beyond the principal axis cranking. This leads to the tilted axis cranking (TAC) which is also called two-dimensional cranking.

The semi-classical mean-field description for tilted axis rotation can be traced back to the 1980s~\cite{Kerman1981Nucl.Phys.,Frisk1987Phys.Lett.}. After the first self-consistent TAC solutions were found in Ref.~\cite{Frauendorf1993Nucl.Phys.}, the qualities of the TAC approximation were discussed and tested in Ref.~\cite{Frauendorf1996ZP} with the particle rotor model (PRM). Because of the high numerical complexity of the TAC model, even today most of the applications are based on simple schematic Hamiltonians, such as the pairing-plus-quadrupole model~\cite{Frauendorf1993Nucl.Phys.}, while fully self-consistent calculations based on universal density functionals are relatively rare. Only recently, relativistic~\cite{Madokoro2000Phys.Rev.C,Peng2008Phys.Rev.C} and non-relativistic~\cite{Olbratowski2002Acta.Phys.Pol.B,Olbratowski2004Phys.Rev.Lett.} density functionals were used for fully microscopic investigations of magnetic rotation and of chirality.

During the past two decades, the covariant density functional theory (CDFT) has received wide attention due to its success in describing many nuclear phenomena in stable as well as exotic nuclei~\cite{Ring1996Prog.Part.Nucl.Phys.,Meng2006Prog.Part.Nucl.Phys.}. On the basis of the same functionals and without any additional parameters, the rotational excitations can be described in the practical applications within the self-consistent cranked relativistic mean-field (RMF) framework. The cranking RMF equations with arbitrary orientation of the rotational axis, i.e., three-dimensional cranking, have been developed in Ref.~\cite{Madokoro2000Phys.Rev.C}. However, because of its numerical complexity, so far, it has been applied only for the magnetic rotation in $^{84}\rm Rb$. Focusing on the magnetic rotation bands, in 2008, a completely new computer code for the self-consistent two-dimensional cranking RMF theory has been established~\cite{Peng2008Phys.Rev.C}. It is based on the non-linear meson-exchange models and includes considerable improvements allowing systematic investigations.

In recent years, RMF models based on point-coupling interactions (RMF-PC)~\cite{Nikolaus1992Phys.Rev.C,Burvenich2002Phys.Rev.C,Zhao2010Phys.Rev.C} have attracted more and more attention owing to the following advantages: firstly, they avoid the use of unphysical mesons with nonlinear self-interactions, especially the fictitious $\sigma$-meson, and the solution of the corresponding Klein-Gordon equations; secondly, it is possible to study the role of naturalness~\cite{Friar1996Phys.Rev.C,Manohar1984Nucl.Phys.} in these effective theories for nuclear-structure-related problems; thirdly, they provide better opportunities to investigate the relationship to nonrelativistic approaches~\cite{Sulaksono2003Ann.Phys.(NY)}; and finally, because of their numerical simplicity it is relatively easy to study effects beyond mean-field for nuclear low-lying collective excited states~\cite{Li2009Phys.Rev.C,Yao2009Phys.Rev.C}.

Therefore, it is worthwhile to develop the tilted cranking RMF model based on a point-coupling interaction. This simplifies the problem considerably. In this work, the self-consistent TAC-RMF model based on a point-coupling interaction will be established. Moreover, for the newly observed shears bands in $^{60}\rm Ni$, microscopic and self-consistent investigations are still missing and strongly desired. Thus, the present model will be applied for a systematic investigation of the shears bands in $^{60}\rm Ni$ with the newly proposed parametrization PC-PK1~\cite{Zhao2010Phys.Rev.C}.

\section{Theoretical framework}

In complete analogy to the successful meson-exchange RMF models~\cite{Ring1996Prog.Part.Nucl.Phys.,Meng2006Prog.Part.Nucl.Phys.},
a RMF-PC model with a zero-range point-coupling interaction is proposed in Ref.~\cite{Nikolaus1992Phys.Rev.C}, i.e., the $\sigma$, $\omega$, and $\rho$ mesons exchange is replaced by the corresponding contact interactions in the scalar-isoscalar, vector-isoscalar, and vector-isovector channels. The starting point of the RMF-PC theory is an effective Lagrangian density of the form
\begin{equation}\label{EQ:LAG}
  {\cal L} = {\cal L}^{\rm free}+{\cal L}^{\rm 4f}+{\cal L}^{\rm hot}+{\cal L}^{\rm der}+{\cal L}^{\rm em},
 \end{equation}
including the Lagrangian density for free nucleons ${\cal L}^{\rm free}$, the four-fermion point-coupling terms ${\cal L}^{\rm 4f}$, the higher order terms ${\cal L}^{\rm hot}$ accounting for the medium effects, the derivative terms ${\cal L}^{\rm der}$ to simulate the effects of finite-range which are crucial for a quantitative description for nuclear density distributions (e.g., nuclear radii), and the electromagnetic interaction terms ${\cal L}^{\rm em}$. The detailed formalism of this RMF-PC model can be seen, e.g., in Ref.~\cite{Zhao2010Phys.Rev.C}.

For applications to magnetic rotation, it is assumed that the nucleus rotates around an axis in the $xz$ plane and the Lagrangian of the RMF-PC model is transformed
into a frame rotating uniformly with a constant rotational frequency,
\begin{equation}
  \bm{\Omega}=(\Omega_x,0,\Omega_z)=(\Omega\cos\theta_\Omega,0,\Omega\sin\theta_\Omega),
\end{equation}
where $\theta_\Omega:=\sphericalangle(\bm{\Omega},\bm{e}_x)$ is the
tilted angle between the cranking axis and the $x$-axis. From this rotating
Lagrangian, the equations of motion can be derived in the same manner as in the meson-exchange case~\cite{Koepf1989Nucl.Phys.,Peng2008Phys.Rev.C} and one finds
 \begin{equation}
   [\bm{\alpha}\cdot(-i\bm{\nabla}-\bm{V})+\beta(m+S)
    +V-\bm{\Omega}\cdot\hat{\bm{J}}]\psi_k=\epsilon_k\psi_k,
 \end{equation}
where $\hat{\bm{J}}=\hat{\bm{L}}+\frac{1}{2}\hat{\bm{\Sigma}}$ is the total angular momentum of the nucleon spinors, and the relativistic fields $S(\bm{r})$, $V^\mu(\bm{r})$ read
 \begin{eqnarray}
   S(\bm{r})&=&\alpha_S\rho_S+\beta_S\rho_S^2+\gamma_S\rho_S^3+\delta_S\triangle\rho_S, \\
   V^\mu(\bm{r})&=&\alpha_V j_V^\mu+\gamma_V(j_V^\mu)^3+\delta_V\triangle j_V^\mu+\tau_3\alpha_{TV} j_{TV}^\mu+\tau_3\delta_{TV}\triangle j_{TV}^\mu+eA^\mu,
 \end{eqnarray}
 with $e$ the electric charge unit vanishing for neutrons. From prior experience in the meson-exchange case~\cite{Koepf1989Nucl.Phys.,Koepf1990Nucl.Phys.}, the spatial components of the electro-magnetic vector potential $\bm{A}(\bm{r})$ are neglected since these contributions are extremely small. However, the derivative terms are kept here since it turns out that their contributions are not negligible. The method used to solve the coupled equations of motion is similar to that used in the case of meson-exchange potentials, however considerably simplified by the fact that one has no Klein-Gordon equations for mesons in the present case.

The total energy in the laboratory frame is given by
  \begin{eqnarray}
      E_{\rm tot}&=&\sum\limits_{k=1}^A\epsilon_k-\int d^3r \left\{\frac{1}{2}\alpha_S\rho_S^2+\frac{1}{2}\alpha_V j_V^\mu (j_V)_\mu\right.\nonumber \\
      &&+\frac{1}{2}\alpha_{TV} j_{TV}^\mu (j_{TV})_\mu+\frac{2}{3}\beta_S\rho_S^3+\frac{3}{4}\gamma_S\rho_S^4\nonumber\\
      &&+\frac{3}{4}\gamma_V(j_V^\mu (j_V)_\mu)^2+\frac{1}{2}\delta_S\rho_S\triangle\rho_S+\frac{1}{2}\delta_V (j_V)_\mu\triangle j_V^\mu\nonumber\\
      &&\left.+\frac{1}{2}\delta_{TV} j_{TV}^\mu\triangle (j_{TV})_\mu
      +\frac{1}{2}e j_p^0A_0
      \right\}+\sum\limits_{k=1}^A\langle k|\bm{\Omega}\hat{\bm{J}}|k\rangle\nonumber\\
      &&+E_{\rm c.m.}.
  \end{eqnarray}
The term of $E_{\rm c.m.}=-\langle\bm{P}^2\rangle/2M$ is the microscopic correction for spurious center-of-mass (c.m.) motion. It has been shown that the microscopic c.m. correction provides more reasonable and reliable results than phenomenological ones~\cite{Bender2000Eur.Phys.J.A,Zhao2009Chin.Phys.Lett.}. Furthermore, following the standard techniques, the root-mean-square (rms) radii, the expectation values of the angular momenta, and the quadrupole moments can be calculated. The detailed formalism and numerical techniques can be seen, e.g., in Ref.~\cite{Peng2008Phys.Rev.C}.

\section{Numerical details}

In this work, the newly proposed parameter set PC-PK1~\cite{Zhao2010Phys.Rev.C} is used, while the pairing correlations are  neglected. The Dirac equation for the nucleons is solved in a three-dimensional harmonic oscillator basis~\cite{Gambhir1990Ann.Phys.(N.Y.)}. By increasing the number $N_f$ of major shells from $N_f=10$ to $N_f=12$, the changes of total energies and total quadrupole moments are within 0.1\% and 4\% for the ground state of the nucleus $^{60}\rm Ni$, respectively. Therefore, a basis of 10 major oscillator shells is adopted in the present calculations.

For the nucleus $^{60}\rm Ni$, four bands denoted as M-1, M-2, M-3, and
M-4 have been reported in Ref.~\cite{Torres2008Phys.Rev.C}, where
it has been suggested that the bands M-1 and M-4 are built from the same type of
configurations, i.e.,
$\pi[(1f_{7/2})^{-1}(fp)^1]\otimes\nu[(1g_{9/2})^1(fp)^3]$, and the
bands M-2 and M-3 are built from the configuration $\pi[(1f_{7/2})^{-1}(1g_{9/2})^1]\otimes\nu[(1g_{9/2})^1(fp)^3]$ or
$\pi[(1f_{7/2})^{-1}(fp)^1]\otimes\nu[(1g_{9/2})^2(fp)^2]$.
In the following, the configurations
$\pi[(1f_{7/2})^{-1}(fp)^1]\otimes\nu[(1g_{9/2})^1(fp)^3]$,
$\pi[(1f_{7/2})^{-1}(1g_{9/2})^1]\otimes\nu[(1g_{9/2})^1(fp)^3]$,
and $\pi[(1f_{7/2})^{-1}(fp)^1]\otimes\nu[(1g_{9/2})^2(fp)^2]$
are referred as Config1, Config2, and Config3, respectively. It should be noted that M-1 and M-4 might be the possible candidates for chiral partner bands~\cite{Frauendorf1997NP}. More detailed investigations in this direction, however, would go beyond two-dimensional cranking and beyond the
scope of the present investigations.

In tilted axis cranking calculations, the tilted angle
$\theta_\Omega$ for a given $\Omega$ is determined in a
self-consistent way either by minimizing the total Routhian
 \begin{equation}
   E'(\Omega, \theta_\Omega)=\langle\hat{H}-\cos\theta_\Omega\Omega\hat{J}_x-\sin\theta_\Omega\Omega\hat{J}_z\rangle,
 \end{equation}
 with respect to the angle $\theta_\Omega$ or by requiring
 that $\bm{\Omega}$ is parallel to $\bm{J}$.
It has been checked that in the present
calculations the direction of the cranking axis $\theta_\Omega$
and the direction of angular momentum $\theta_J$ are identical, which means that
self-consistency has been fully achieved.

\section{Results and discussion}

\begin{figure*}[!htbp]
\centering
\includegraphics[width=12cm]{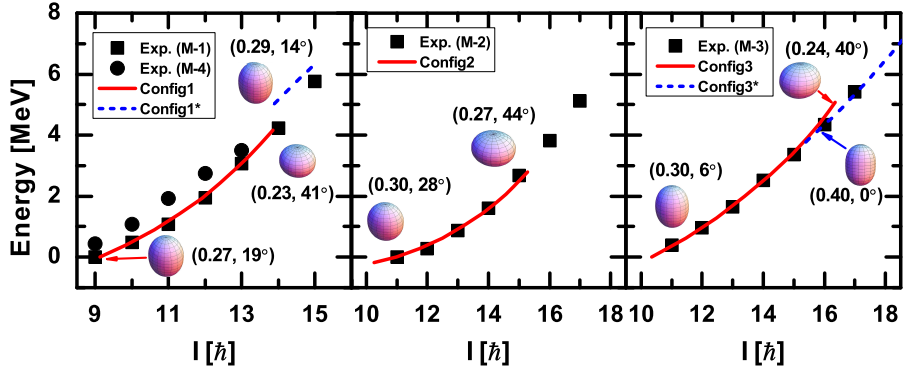}
\caption{(Color online) Energy spectra
obtained from the TAC-RMF calculations in comparison with the available data for bands M-1 and
M-4 (left panel), M-2 (middle panel), as well as M-3 (right panel).
The energies at $I=9\hbar$, $I=11\hbar$, and $I=15\hbar$ are taken
as references in the left, middle and right panels, respectively. The evolutions of the nuclear shape $(\beta,\gamma)$ for bands M-1, M-2, and M-3 are also illustrated with the schematic pictures.}
\label{fig:energy}
\end{figure*}

In Fig.~\ref{fig:energy} the calculated energy spectra are compared with the available data for the bands M-1 and M-4 (left
panel), M-2 (middle panel), as well as M-3 (right panel). The experimental energies of the bands M-1, M-2, and M-3 are
reproduced in general very well by these TAC-RMF calculations.
However, the assigned configuration for each of these bands could not be followed in
the calculations up to the largest observed spin values, i.e., convergent
results could be obtained only up to $\sim14\hbar$ for Config1,
$\sim15\hbar$ for Config2, and $\sim16\hbar$ for Config3.

By increasing the rotational frequency, it is found that the configurations, $\pi[(1f_{7/2})^{-1}(fp)^1]\otimes\nu[(1g_{9/2})^{1}(fp)^4(1f_{7/2})^{-1}]$ (Config1*) and $\pi[(1f_{7/2})^{-2}(fp)^2]\otimes\nu[(1g_{9/2})^{2}(fp)^3(1f_{7/2})^{-1}]$ (Config3*) compete strongly with Config1 and Config3, respectively. One finds that a
pair of neutrons in the $f_{7/2}$ shell is broken at $I=15\hbar$ in band M-1. In the band M-3, one observes for $I=16\hbar$ that apart from a broken pair of neutrons in the $f_{7/2}$ shell, a unpaired proton in the $f_{7/2}$ shell moves to the $fp$ orbital.

The shape evolutions of bands M-1, M-2, and M-3 are also shown in Fig.~\ref{fig:energy}.
It is interesting to note that the nucleus changes its shape from prolate-like to oblate-like with the increasing frequency in Config1, Config2, and Config3, and comes back to a prolate-like deformation with the configuration changing from Config1 to Config1*, and Config3 to Config3*. In particular, the nucleus with Config3* has a relatively large deformation ($\beta\sim0.4$) with axial symmetry.

\begin{figure*}[!htbp]
\centering
\includegraphics[width=12cm]{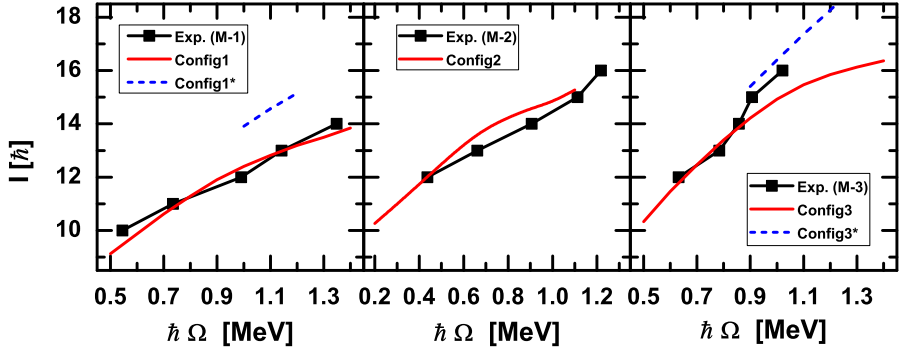}
\caption{(Color online) Total angular momenta as functions of the rotational frequency in the TAC-RMF calculations in comparison with the data for bands M-1 (left panel), M-2 (middle panel), and M-3 (right panel).}
\label{fig:spin}
\end{figure*}

Fig.~\ref{fig:spin} shows a comparison between the experimental and calculated total angular momenta as functions of the rotational frequency for the bands M-1 (left panel), M-2 (middle panel), and M-3 (right panel). The experimental rotational frequency can be extracted as in Ref.~\cite{Frauendorf1996ZP}
\begin{equation}
  \hbar\Omega_{\rm exp}=\frac{1}{2}[E_{\gamma}(I+1\rightarrow I)+E_{\gamma}(I\rightarrow I-1)]\approx\frac{dE}{dI}.
\end{equation}
It is found that the calculated total angular momenta agree well with the data. This indicates that the present TAC-RMF calculations can reproduce the moments of inertia rather well.
The same good agreement has been found also in relativistic TAC calculations for the nucleus $^{142}$Gd in Ref.~\cite{Peng2008Phys.Rev.C}, where also non-relativistic Skyrme calculations are available~\cite{Olbratowski2002Acta.Phys.Pol.B}. However, in these non-relativistic calculations the theoretical moments of inertia deviate considerably from experiment. The good agreement in the relativistic models is obviously connected with the fact that in CDFT, because of Lorentz invariance, time-even and time-odd mean fields are coupled with the same coupling constants, whereas in many non-relativistic calculations the coupling constants of the time-odd parts are not adjusted properly. Of course pairing correlations have been neglected in all realistic TAC calculations available at present, relativistic and non-relativistic. This may have consequences and will be investigated in
near future.

At $I=15\hbar$, a band crossing is observed experimentally in both the M-2 and M-3 bands. It corresponds to a change of the configuration. Remarkably, the observed band crossing is excellently reproduced by the calculation with Config3 and Config3*. Further experiments are welcome to validate the predicted band crossing for the band M-1 related to Config1*, especially for the data at $I=15\hbar$ in the band M-1.

\begin{figure*}[!htbp]
\centering
\includegraphics[width=12cm]{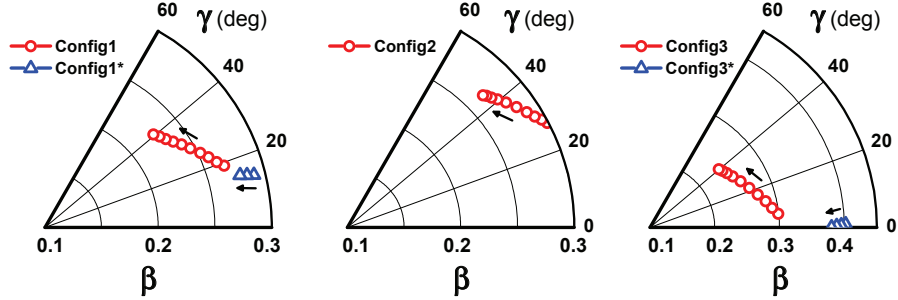}
\caption{(Color online) Evolutions of deformation parameters $\beta$ and $\gamma$ driven by increasing rotational frequency in the TAC-RMF calculations for the M-1 (left panel), M-2 (middle panel) and M-3 (right panel) bands. The arrows indicate the increasing direction of rotational frequency.}
\label{fig:deformation}
\end{figure*}

As shown in Fig.~\ref{fig:energy}, the nucleus changes its shape when the rotational frequency increases.
This can be seen more explicitly in Fig.~\ref{fig:deformation}, which shows the evolutions of deformation parameters $\beta$ and $\gamma$ driven by increasing rotational frequency in the TAC-RMF calculations.
At the bandhead, the deformation parameters ($\beta$, $\gamma$) are ($0.27,19^\circ$) for M-1, ($0.30,28^\circ$) for M-2, and ($0.30,6^\circ$) for M-3. With the increase of rotational frequency, the nucleus becomes oblate gradually.
In detail, the $\beta$ values for Config1, Config2, and Config3, typically lying between 0.2 and 0.3, decrease smoothly with the rotational frequency $\Omega$, while the $\gamma$ values show a smoothly increasing tendency up to about $40^\circ$.
For the Config1* and Config3*, with one broken pair of neutron in the $f_{7/2}$ orbital, the $\beta$ values jump from $0.23$ to $0.29$ and from $0.24$ to $0.40$, respectively, and decrease again with the rotational frequency $\Omega$. Meanwhile, the $\gamma$ value for Config1* decreases from $41^\circ$ to $14^\circ$ and that for Config3* reduces from $40^\circ$ to nearly zero. Again we note that the $\gamma$-deformations for configurations Config1-3 are in accordance with a possible appearance of chiral partner bands~\cite{Frauendorf1997NP}. This will be interesting in a future investigation.

\begin{figure}[!htbp]
\centering
\includegraphics[width=8cm]{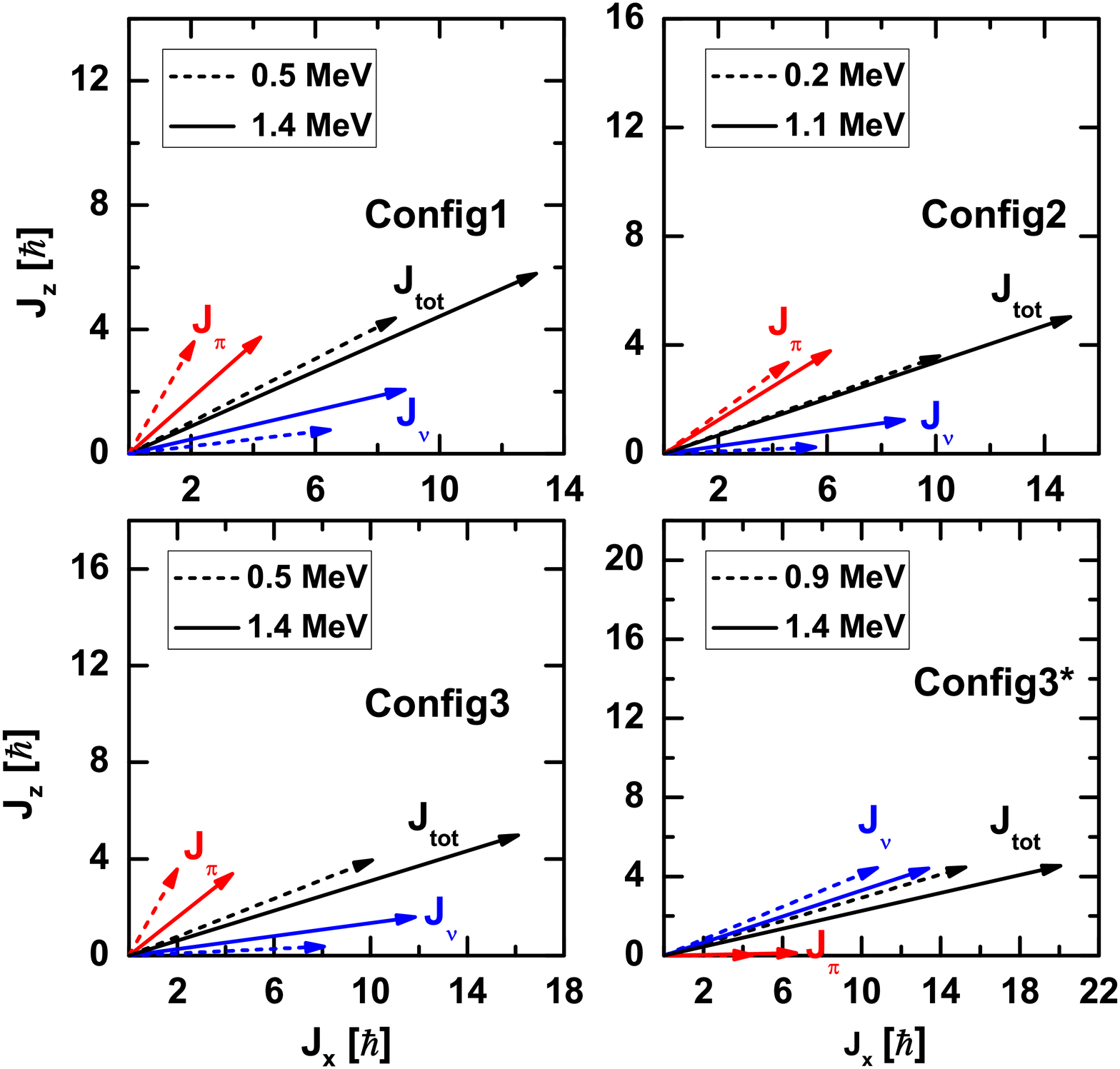}
\caption{(Color online) Composition of the total angular momentum at both the bandhead and the maximum rotational frequency in the TAC-RMF calculations with the configurations of Config1, Config2, Config3, and Config3*.}
\label{fig:vector}
\end{figure}

In order to examine the shears mechanism for the magnetic rotation bands in $^{60}\rm Ni$, the proton and neutron angular momentum vectors $\bm{J}_\pi$ and $\bm{J}_\nu$ as well as the total angular momentum vector $\bm{J}_{\rm tot}=\bm{J}_\pi+\bm{J}_\nu$ at both the bandhead and the maximum rotational frequency in the TAC-RMF calculations with the configurations of Config1, Config2, Config3, and Config3* are shown in Fig.~\ref{fig:vector}. The proton and neutron angular momenta $\bm{J}_\pi$ and $\bm{J}_\nu$ are defined as
 \begin{equation}
   \bm{J}_\pi=\langle\hat{\bm{J}}_\pi\rangle=\sum\limits_{p=1}^Z\langle p|\hat J|p\rangle,\quad
   \bm{J}_\nu=\langle\hat{\bm{J}}_\nu\rangle=\sum\limits_{n=1}^N\langle n|\hat J|n\rangle,
 \end{equation}
where the sum runs over all the proton (or neutron) levels occupied in the cranking wave function in the intrinsic system.

For the bands built on Config1, Config2, and Config3, the contributions to the angular momenta come mainly from the high $j$ orbitals, i.e., the $g_{9/2}$ neutron(s) and the $f_{7/2}$ proton. At the bandhead, the neutron particle(s) filling the bottom of the $g_{9/2}$ shell mainly contribute to the neutron angular momentum along the
$z$-axis, and the proton hole at the upper end of the $f_{7/2}$ shell mainly contributes to the proton angular momentum along the $x$-axis. They form the two blades of the shears. As the frequency increases, the two blades move toward each other to provide larger angular momentum, while the direction of the total angular momentum stays nearly unchanged. In this way, the shears mechanism is clearly seen.

One should notice that the proton particle in the $g_{9/2}$ orbital also give substantial contributions to the proton angular momentum in the case of Config2. As a result, $\bm{J}_\pi$ has not only a large $J_z$ component but also a substantial $J_x$ component even at the bandhead. Thus, the shears angle $\Theta$, the angle between $\bm{J}_\pi$ and $\bm{J}_\nu$, is not as large as those of Config1 and Config3, and decreases only by a small amount with increasing rotational frequency.

For the Config3*, as the two proton holes in the $f_{7/2}$ orbital are paired, the proton angular momentum comes mainly from the particles in the $fp$ shell, which aligns along the $x$-axis. The neutron hole in the $f_{7/2}$ orbital gives substantial contributions to the neutron angular momentum, which leads to a large $J_z$ component. Higher spin states in the band are created by aligning the neutron angular momentum towards the $x$-axis.
Considering the large axially symmetric prolate deformation as shown in Fig.~\ref{fig:deformation}, the mechanism of producing higher spin states with Config3* is electric rotation rather than magnetic rotation. Therefore, we observe a transition from magnetic rotation to electric rotation in Config3*.

\begin{figure*}[!htbp]
\centering
\includegraphics[width=10cm]{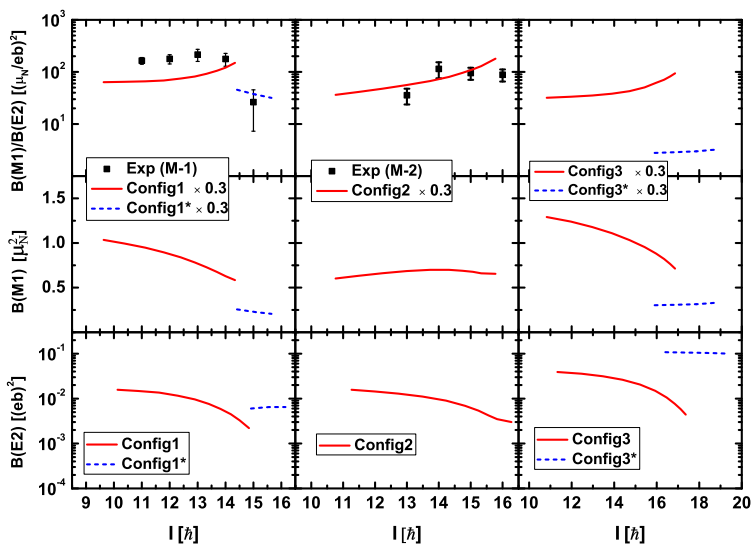}
\caption{(Color online) $B(M1)/B(E2)$ ratios, reduced $M1$ transition probabilities $B(M1)$, and reduced $E2$ transition probabilities $B(E2)$ for the M-1, M-2 and M-3 bands as functions of the total angular momentum in the TAC-RMF calculations in comparison with the available data.}
\label{fig:MEtransi}
\end{figure*}

Typical characteristics of magnetic rotation are the strongly enhanced $M1$ transitions at low angular momenta as well as their decreasing tendency with the increasing angular momentum. In contrast, the $E2$ transitions of the band are very weak. In Fig.~\ref{fig:MEtransi}, the calculated $B(M1)/B(E2)$ ratios, reduced $M1$ transition probabilities $B(M1)$, and reduced $E2$ transition probabilities $B(E2)$ for the M-1, M-2, and M-3 bands are shown as functions of the total angular momentum in the TAC-RMF calculations in comparison with the available data. Here the $B(M1)$ values are derived from the relativistic expression of the effective current operator~\cite{Peng2008Phys.Rev.C}  As in Ref.~\cite{Madokoro2000Phys.Rev.C} they are attenuated by the same factor 0.3. At the moment there is no microscopic derivation of this factor 0.3. There are several reasons for the deviations of the observed $B(M1)/B(E2)$ ratios from the pure mean field values and none of them is taken into account in the present calculations: (a) Pairing correlations strongly affect the levels in the neighborhood of the Fermi surface. This causes a stronger reduction for the B(M1) values with major contributions from the valence particles or holes ($\Delta\ell=0$) than for the B(E2) values with large contributions from the core ($\Delta\ell=2$). (b) The coupling to complex configurations such as particle-vibration coupling (Arima-Horie effect~\cite{Arima1954Prog.Theor.Phys.509,Arima2011SciChinaSerG-PhysMechAstron188}) leads in all cases to a quenching of the $B(M1)$-values for neutron configurations~\cite{Bauer1973Nucl.Phys.A535,Matsuzaki1988Prog.Theor.Phys.836} and an increase of the $B(E2)$-values through effective charges. (c) Meson exchange currents and higher corrections  also cause a reduction of the effective $g$-factors for the neutrons~\cite{Towner1987Phys.Rep.263,Li2010,Li2011}.

However, it is not the absolute value of the $B(M1)/B(E2)$ ratios, which characterizes the shear bands, but rather the behavior of these values with increasing angular momentum.
 As seen from Fig.~\ref{fig:MEtransi}, this behavior is in good agreement with the data for the bands M-1 and M-2. In particular, the high-spin states in the band M-1 ($I=14\hbar, 15\hbar$) can be reproduced in the calculations with the Config1*. Meanwhile, the predicted tendency of the $B(M1)/B(E2)$ ratios for Config3 is similar to those for Config1 and Config2. In addition, the $B(M1)/B(E2)$ ratios for Config3* are one order of magnitude smaller than those for Config3, which might result from the transitions from magnetic rotation in Config3 to electric rotation in Config3*.

For the $B(M1)$ values, the smooth-decreasing tendency shown in bands M-1 and M-3 obviously presents the shears mechanism. Owing to the smaller shears angles $\Theta$ of Config1* and Config3* in comparison with those of Config1 and Config3, the $B(M1)$ values drop suddenly with the change of configurations. For the band M-2, the present work predicts nearly constant $B(M1)$ values with increasing spin, which corresponds to the small decline of the shears angle $\Theta$ of Config2 as shown in Fig.~\ref{fig:vector}.

In contrast to the large $B(M1)$ values (about several $\mu_N^2$), the $B(E2)$ values are very small for all the three bands ($<0.1~(eb)^2$). Moreover, with increasing spin, a sharp decreasing trend of the $B(E2)$ values can be found from the calculated results with Config1, Config2, and Config3. This indicates a signal for the termination of the corresponding shears band. It is noted that the termination of band M-1 has been pointed out in Ref.~\cite{Torres2008Phys.Rev.C} by analyzing the coupling of the spins of valence nucleons. When the configuration of band M-1 (M-3) changes from Config1 (Config3) to Config1* (Config3*) due to the broken of nucleon pairs, the corresponding $B(E2)$ values are predicted as 0.006 (0.1)~$(eb)^2$, which is of the same order of magnitude as the $B(E2)$ value at the bandhead of Config1 (Config3) as shown in Fig.~\ref{fig:MEtransi}.

\section{Summary}

In summary, the self-consistent TAC-RMF theory based on a point-coupling interaction has been established and applied to investigate the newly observed shears bands in $^{60}\rm Ni$ systematically. The tilted angles, deformation parameters, energy spectra, and reduced $M1$ and $E2$ transition probabilities have been calculated in a fully microscopic and self-consistent way for various configurations and rotational frequencies.

It is found that the experimental energies of bands M-1, M-2, and M-3 are reproduced very well by these TAC-RMF calculations. The comparison of experimental and calculated angular momenta as a function of frequency indicates a clear evidence for a band crossing. Of special interest is the strong competition between configurations Config1* (Config3*) and Config1 (Config3) in the bands M-1 and M-3. In the diagrams showing the contributions to the angular momentum, it becomes evident that the mechanism of producing higher spin states in the band M-3 is the transition from the magnetic rotation in Config3 to the electric rotation in Config3* which is connected with a large prolate deformation ($\beta\sim0.4$).

The tendency of experimental $B(M1)/B(E2)$ ratios for bands M-1 and M-2 is in a good agreement with the data. The fact that the $B(M1)$ values decrease with the increasing frequency in bands M-1 and M-3 as well as the fact that the $B(E2)$ values are very small ($<0.1~(eb)^2$) indicates the appearance of the shears mechanism in $^{60} \rm Ni$.
Moreover, with increasing spin, a sharp decreasing trend of the $B(E2)$ values can be found for the calculated results for Config1, Config2, and Config3, which indicates the termination of the corresponding shears bands.

\section*{Acknowledgements}
This work was partly supported by the Major State 973 Program
2007CB815000, the National Natural Science Foundation of China under
Grants No. 10975007, No. 10975008, and No. 11005069, China
Postdoctoral Science Foundation Grant No. 20100480149, and the DFG
cluster of excellence \textquotedblleft Origin and Structure of the
Universe\textquotedblright\ (www.universe-cluster.de).

\newpage

\begin{thebibliography}{44}
\expandafter\ifx\csname natexlab\endcsname\relax\def\natexlab#1{#1}\fi
\providecommand{\bibinfo}[2]{#2}
\ifx\xfnm\relax \def\xfnm[#1]{\unskip,\space#1}\fi
\bibitem[{Johnson et~al.(1971)Johnson, Ryde, and
  Sztarkier}]{Johnson1971Phys.Lett.}
\bibinfo{author}{A.~Johnson}, \bibinfo{author}{H.~Ryde},
  \bibinfo{author}{J.~Sztarkier}, \bibinfo{journal}{Phys. Lett.}
  \bibinfo{volume}{B34} (\bibinfo{year}{1971}) \bibinfo{pages}{605}.
\bibitem[{Stephens and Simon(1972)}]{Stephens1972Nucl.Phys.}
\bibinfo{author}{F.~S. Stephens}, \bibinfo{author}{R.~S. Simon},
  \bibinfo{journal}{Nucl. Phys.} \bibinfo{volume}{A183} (\bibinfo{year}{1972})
  \bibinfo{pages}{257}.
\bibitem[{Banerjee et~al.(1973)Banerjee, Mang, and
  Ring}]{Banerjee1973Nucl.Phys.}
\bibinfo{author}{B.~Banerjee}, \bibinfo{author}{H.~J. Mang},
  \bibinfo{author}{P.~Ring}, \bibinfo{journal}{Nucl. Phys.}
  \bibinfo{volume}{A215} (\bibinfo{year}{1973}) \bibinfo{pages}{366}.
\bibitem[{Twin and et~al.(1986)}]{Twin1986Phys.Rev.Lett.}
\bibinfo{author}{P.~J. Twin}, \bibinfo{author}{et~al.}, \bibinfo{journal}{Phys.
  Rev. Lett.} \bibinfo{volume}{57} (\bibinfo{year}{1986}) \bibinfo{pages}{811}.
\bibitem[{Frauendorf et~al.(1994)Frauendorf, Meng, and Reif}]{Frauendorf1994}
\bibinfo{author}{S.~Frauendorf}, \bibinfo{author}{J.~Meng},
  \bibinfo{author}{J.~Reif}, in: \bibinfo{editor}{M.~A. Deleplanque} (Ed.),
  \bibinfo{booktitle}{Proceedings of the Conference on Physics From Large
  $\gamma$-Ray Detector Arrays}, volume \bibinfo{volume}{II of Report
  LBL35687}, \bibinfo{publisher}{Univ. of California, Berkeley},
  \bibinfo{year}{1994}, p.~\bibinfo{pages}{52}.
\bibitem[{Frauendorf(1997)}]{Frauendorf1997Z.Phys.A}
\bibinfo{author}{S.~Frauendorf}, \bibinfo{journal}{Z. Phys. A}
  \bibinfo{volume}{358} (\bibinfo{year}{1997}) \bibinfo{pages}{163}.
\bibitem[{Frauendorf and Meng(1997)}]{Frauendorf1997NP}
\bibinfo{author}{S.~Frauendorf}, \bibinfo{author}{J.~Meng},
  \bibinfo{journal}{Nucl. Phys.} \bibinfo{volume}{A617} (\bibinfo{year}{1997})
  \bibinfo{pages}{131}.
\bibitem[{Frauendorf(1993)}]{Frauendorf1993Nucl.Phys.}
\bibinfo{author}{S.~Frauendorf}, \bibinfo{journal}{Nucl. Phys.}
  \bibinfo{volume}{A557} (\bibinfo{year}{1993}) \bibinfo{pages}{259c}.
\bibitem[{Clark and et~al.(1992)}]{Clark1992Phys.Lett.}
\bibinfo{author}{R.~Clark}, \bibinfo{author}{et~al.}, \bibinfo{journal}{Phys.
  Lett.} \bibinfo{volume}{B275} (\bibinfo{year}{1992}) \bibinfo{pages}{247}.
\bibitem[{Baldsiefen and et~al.(1992)}]{Baldsiefen1992Phys.Lett.}
\bibinfo{author}{G.~Baldsiefen}, \bibinfo{author}{et~al.},
  \bibinfo{journal}{Phys. Lett.} \bibinfo{volume}{B275} (\bibinfo{year}{1992})
  \bibinfo{pages}{252}.
\bibitem[{Kuhnert and et~al.(1992)}]{Kuhnert1992Phys.Rev.C}
\bibinfo{author}{A.~Kuhnert}, \bibinfo{author}{et~al.}, \bibinfo{journal}{Phys.
  Rev. C} \bibinfo{volume}{46} (\bibinfo{year}{1992}) \bibinfo{pages}{133}.
\bibitem[{Clark and et~al.(1997)}]{Clark1997Phys.Rev.Lett.}
\bibinfo{author}{R.~M. Clark}, \bibinfo{author}{et~al.},
  \bibinfo{journal}{Phys. Rev. Lett.} \bibinfo{volume}{78}
  (\bibinfo{year}{1997}) \bibinfo{pages}{1868}.
\bibitem[{Amita et~al.(2000)Amita, Jain, and
  Singh}]{Amita2000AtomicDataandNuclearDataTables}
\bibinfo{author}{Amita}, \bibinfo{author}{A.~K. Jain},
  \bibinfo{author}{B.~Singh}, \bibinfo{journal}{At. Data Nucl. Data Tables}
  \bibinfo{volume}{74} (\bibinfo{year}{2000}) \bibinfo{pages}{283}.
  \bibinfo{note}{Revised edition at
  [http://www.nndc.bnl.gov/publications/preprints/mag-dip-rot-bands.pdf]}.
\bibitem[{Torres and et~al.(2008)}]{Torres2008Phys.Rev.C}
\bibinfo{author}{D.~A. Torres}, \bibinfo{author}{et~al.},
  \bibinfo{journal}{Phys. Rev. C} \bibinfo{volume}{78} (\bibinfo{year}{2008})
  \bibinfo{pages}{054318}.
\bibitem[{Frauendorf et~al.(1996)Frauendorf, Reif, and
  Winter}]{Frauendorf1996Nucl.Phys.}
\bibinfo{author}{S.~Frauendorf}, \bibinfo{author}{J.~Reif},
  \bibinfo{author}{G.~Winter}, \bibinfo{journal}{Nucl. Phys.}
  \bibinfo{volume}{A601} (\bibinfo{year}{1996}) \bibinfo{pages}{41}.
\bibitem[{Kerman and Naoki(1981)}]{Kerman1981Nucl.Phys.}
\bibinfo{author}{A.~K. Kerman}, \bibinfo{author}{O.~Naoki},
  \bibinfo{journal}{Nucl. Phys.} \bibinfo{volume}{A361} (\bibinfo{year}{1981})
  \bibinfo{pages}{179}.
\bibitem[{Frisk and Bengtsson(1987)}]{Frisk1987Phys.Lett.}
\bibinfo{author}{H.~Frisk}, \bibinfo{author}{R.~Bengtsson},
  \bibinfo{journal}{Phys. Lett.} \bibinfo{volume}{B196} (\bibinfo{year}{1987})
  \bibinfo{pages}{14}.
\bibitem[{Frauendorf and Meng(1996)}]{Frauendorf1996ZP}
\bibinfo{author}{S.~Frauendorf}, \bibinfo{author}{J.~Meng},
  \bibinfo{journal}{Z. Phys. A} \bibinfo{volume}{356} (\bibinfo{year}{1996})
  \bibinfo{pages}{263}.
\bibitem[{Madokoro et~al.(2000)Madokoro, Meng, Matsuzaki, and
  Yamaji}]{Madokoro2000Phys.Rev.C}
\bibinfo{author}{H.~Madokoro}, \bibinfo{author}{J.~Meng},
  \bibinfo{author}{M.~Matsuzaki}, \bibinfo{author}{S.~Yamaji},
  \bibinfo{journal}{Phys. Rev. C} \bibinfo{volume}{62} (\bibinfo{year}{2000})
  \bibinfo{pages}{061301}.
\bibitem[{Peng et~al.(2008)Peng, Meng, Ring, and Zhang}]{Peng2008Phys.Rev.C}
\bibinfo{author}{J.~Peng}, \bibinfo{author}{J.~Meng},
  \bibinfo{author}{P.~Ring}, \bibinfo{author}{S.~Q. Zhang},
  \bibinfo{journal}{Phys. Rev. C} \bibinfo{volume}{78} (\bibinfo{year}{2008})
  \bibinfo{pages}{024313}.
\bibitem[{Olbratowski et~al.(2002)Olbratowski, Dobaczewski, Dudek, Rzaca-Urban,
  Marcinkowska, and Lieder}]{Olbratowski2002Acta.Phys.Pol.B}
\bibinfo{author}{P.~Olbratowski}, \bibinfo{author}{J.~Dobaczewski},
  \bibinfo{author}{J.~Dudek}, \bibinfo{author}{T.~Rzaca-Urban},
  \bibinfo{author}{Z.~Marcinkowska}, \bibinfo{author}{R.~M. Lieder},
  \bibinfo{journal}{Acta Phys. Pol. B} \bibinfo{volume}{33}
  (\bibinfo{year}{2002}) \bibinfo{pages}{389}.
\bibitem[{Olbratowski et~al.(2004)Olbratowski, Dobaczewski, Dudek, and
  P\l{}\'ociennik}]{Olbratowski2004Phys.Rev.Lett.}
\bibinfo{author}{P.~Olbratowski}, \bibinfo{author}{J.~Dobaczewski},
  \bibinfo{author}{J.~Dudek}, \bibinfo{author}{W.~P\l{}\'ociennik},
  \bibinfo{journal}{Phys. Rev. Lett.} \bibinfo{volume}{93}
  (\bibinfo{year}{2004}) \bibinfo{pages}{052501}.
\bibitem[{Ring(1996)}]{Ring1996Prog.Part.Nucl.Phys.}
\bibinfo{author}{P.~Ring}, \bibinfo{journal}{Prog. Part. Nucl. Phys.}
  \bibinfo{volume}{37} (\bibinfo{year}{1996}) \bibinfo{pages}{193}.
\bibitem[{Meng et~al.(2006)Meng, Toki, Zhou, Zhang, Long, and
  Geng}]{Meng2006Prog.Part.Nucl.Phys.}
\bibinfo{author}{J.~Meng}, \bibinfo{author}{H.~Toki},
  \bibinfo{author}{S.~Zhou}, \bibinfo{author}{S.~Zhang},
  \bibinfo{author}{W.~Long}, \bibinfo{author}{L.~Geng}, \bibinfo{journal}{Prog.
  Part. Nucl. Phys.} \bibinfo{volume}{57} (\bibinfo{year}{2006})
  \bibinfo{pages}{470}.
\bibitem[{Nikolaus et~al.(1992)Nikolaus, Hoch, and
  Madland}]{Nikolaus1992Phys.Rev.C}
\bibinfo{author}{B.~A. Nikolaus}, \bibinfo{author}{T.~Hoch},
  \bibinfo{author}{D.~G. Madland}, \bibinfo{journal}{Phys. Rev. C}
  \bibinfo{volume}{46} (\bibinfo{year}{1992}) \bibinfo{pages}{1757}.
\bibitem[{B\"urvenich et~al.(2002)B\"urvenich, Madland, Maruhn, and
  Reinhard}]{Burvenich2002Phys.Rev.C}
\bibinfo{author}{T.~B\"urvenich}, \bibinfo{author}{D.~G. Madland},
  \bibinfo{author}{J.~A. Maruhn}, \bibinfo{author}{P.-G. Reinhard},
  \bibinfo{journal}{Phys. Rev. C} \bibinfo{volume}{65} (\bibinfo{year}{2002})
  \bibinfo{pages}{044308}.
\bibitem[{Zhao et~al.(2010)Zhao, Li, Yao, and Meng}]{Zhao2010Phys.Rev.C}
\bibinfo{author}{P.~W. Zhao}, \bibinfo{author}{Z.~P. Li},
  \bibinfo{author}{J.~M. Yao}, \bibinfo{author}{J.~Meng},
  \bibinfo{journal}{Phys. Rev. C} \bibinfo{volume}{82} (\bibinfo{year}{2010})
  \bibinfo{pages}{054319}.
\bibitem[{Friar et~al.(1996)Friar, Madland, and Lynn}]{Friar1996Phys.Rev.C}
\bibinfo{author}{J.~L. Friar}, \bibinfo{author}{D.~G. Madland},
  \bibinfo{author}{B.~W. Lynn}, \bibinfo{journal}{Phys. Rev. C}
  \bibinfo{volume}{53} (\bibinfo{year}{1996}) \bibinfo{pages}{3085}.
\bibitem[{Manohar and Georgi(1984)}]{Manohar1984Nucl.Phys.}
\bibinfo{author}{A.~Manohar}, \bibinfo{author}{H.~Georgi},
  \bibinfo{journal}{Nucl. Phys.} \bibinfo{volume}{B 234} (\bibinfo{year}{1984})
  \bibinfo{pages}{189}.
\bibitem[{Sulaksono et~al.(2003)Sulaksono, B\"urvenich, Maruhn, Reinhard, and
  Greiner}]{Sulaksono2003Ann.Phys.(NY)}
\bibinfo{author}{A.~Sulaksono}, \bibinfo{author}{T.~B\"urvenich},
  \bibinfo{author}{J.~A. Maruhn}, \bibinfo{author}{P.~G. Reinhard},
  \bibinfo{author}{W.~Greiner}, \bibinfo{journal}{Ann. Phys. (NY)}
  \bibinfo{volume}{308} (\bibinfo{year}{2003}) \bibinfo{pages}{354}.
\bibitem[{Li et~al.(2009)Li, Nik\v{s}i\'c, Vretenar, Meng, Lalazissis, and
  Ring}]{Li2009Phys.Rev.C}
\bibinfo{author}{Z.~P. Li}, \bibinfo{author}{T.~Nik\v{s}i\'c},
  \bibinfo{author}{D.~Vretenar}, \bibinfo{author}{J.~Meng},
  \bibinfo{author}{G.~A. Lalazissis}, \bibinfo{author}{P.~Ring},
  \bibinfo{journal}{Phys. Rev. C} \bibinfo{volume}{79} (\bibinfo{year}{2009})
  \bibinfo{pages}{054301}.
\bibitem[{Yao et~al.(2009)Yao, Meng, Ring, and
  Pena~Arteaga}]{Yao2009Phys.Rev.C}
\bibinfo{author}{J.~M. Yao}, \bibinfo{author}{J.~Meng},
  \bibinfo{author}{P.~Ring}, \bibinfo{author}{D.~Pena~Arteaga},
  \bibinfo{journal}{Phys. Rev. C} \bibinfo{volume}{79} (\bibinfo{year}{2009})
  \bibinfo{pages}{044312}.
\bibitem[{Koepf and Ring(1989)}]{Koepf1989Nucl.Phys.}
\bibinfo{author}{W.~Koepf}, \bibinfo{author}{P.~Ring}, \bibinfo{journal}{Nucl.
  Phys.} \bibinfo{volume}{A493} (\bibinfo{year}{1989}) \bibinfo{pages}{61}.
\bibitem[{Koepf and Ring(1990)}]{Koepf1990Nucl.Phys.}
\bibinfo{author}{W.~Koepf}, \bibinfo{author}{P.~Ring}, \bibinfo{journal}{Nucl.
  Phys.} \bibinfo{volume}{A511} (\bibinfo{year}{1990}) \bibinfo{pages}{279}.
\bibitem[{Bender et~al.(2000)Bender, Rutz, Reinhard, and
  Maruhn}]{Bender2000Eur.Phys.J.A}
\bibinfo{author}{M.~Bender}, \bibinfo{author}{K.~Rutz}, \bibinfo{author}{P.-G.
  Reinhard}, \bibinfo{author}{J.~Maruhn}, \bibinfo{journal}{Eur. Phys. J. A}
  \bibinfo{volume}{7} (\bibinfo{year}{2000}) \bibinfo{pages}{467}.
\bibitem[{Zhao et~al.(2009)Zhao, Sun, and Meng}]{Zhao2009Chin.Phys.Lett.}
\bibinfo{author}{P.~Zhao}, \bibinfo{author}{B.~Sun}, \bibinfo{author}{J.~Meng},
  \bibinfo{journal}{Chin. Phys. Lett.} \bibinfo{volume}{26}
  (\bibinfo{year}{2009}) \bibinfo{pages}{112102}.
\bibitem[{Gambhir et~al.(1990)Gambhir, Ring, and
  Thimet}]{Gambhir1990Ann.Phys.(N.Y.)}
\bibinfo{author}{Y.~K. Gambhir}, \bibinfo{author}{P.~Ring},
  \bibinfo{author}{A.~Thimet}, \bibinfo{journal}{Ann. Phys. (N.Y.)}
  \bibinfo{volume}{198} (\bibinfo{year}{1990}) \bibinfo{pages}{132}.
\bibitem[{Arima and Horie(1954)}]{Arima1954Prog.Theor.Phys.509}
\bibinfo{author}{A.~Arima}, \bibinfo{author}{H.~Horie}, \bibinfo{journal}{Prog.
  Theor. Phys.} \bibinfo{volume}{11} (\bibinfo{year}{1954})
  \bibinfo{pages}{509}.
\bibitem[{Arima(2011)}]{Arima2011SciChinaSerG-PhysMechAstron188}
\bibinfo{author}{A.~Arima}, \bibinfo{journal}{Sci China Ser G-Phys Mech Astron}
  \bibinfo{volume}{54} (\bibinfo{year}{2011}) \bibinfo{pages}{188}.
\bibitem[{Bauer et~al.(1973)Bauer, Speth, Klemt, Ring, Werner, and
  Yamazaki}]{Bauer1973Nucl.Phys.A535}
\bibinfo{author}{R.~Bauer}, \bibinfo{author}{J.~Speth},
  \bibinfo{author}{V.~Klemt}, \bibinfo{author}{P.~Ring},
  \bibinfo{author}{E.~Werner}, \bibinfo{author}{T.~Yamazaki},
  \bibinfo{journal}{Nucl. Phys. A} \bibinfo{volume}{209} (\bibinfo{year}{1973})
  \bibinfo{pages}{535}.
\bibitem[{Matsuzaki et~al.(1988)Matsuzaki, Shimizu, and
  Matsuyanagi}]{Matsuzaki1988Prog.Theor.Phys.836}
\bibinfo{author}{M.~Matsuzaki}, \bibinfo{author}{Y.~R. Shimizu},
  \bibinfo{author}{K.~Matsuyanagi}, \bibinfo{journal}{Prog. Theor. Phys.}
  \bibinfo{volume}{79} (\bibinfo{year}{1988}) \bibinfo{pages}{836}.
\bibitem[{Towner(1987)}]{Towner1987Phys.Rep.263}
\bibinfo{author}{I.~S. Towner}, \bibinfo{journal}{Phys. Rep.}
  \bibinfo{volume}{155} (\bibinfo{year}{1987}) \bibinfo{pages}{263}.
\bibitem[{Li(2010)}]{Li2010}
\bibinfo{author}{J.~Li}, \bibinfo{author}{J. M. Yao}, \bibinfo{author}{J. Meng}, \bibinfo{author}{A. Arima},
\bibinfo{journal}{arXiv:1002.0107[nucl-th]}.
\bibitem[{Li(2011)}]{Li2011}
\bibinfo{author}{J.~Li}, \bibinfo{author}{J. Meng}, \bibinfo{author}{P. Ring},
\bibinfo{author}{J. M. Yao}, \bibinfo{author}{A. Arima},
\bibinfo{journal}{Sci China Ser G-Phys Mech Astron}
\bibinfo{volume}{54} (\bibinfo{year}{2011}) \bibinfo{pages}{204}.
\end{thebibliography}

\end{document}